\documentclass[%
 aip,
 jmp,%
 amsmath,amssymb,
reprint,
]{revtex4-1}

\usepackage{graphicx}
\usepackage{dcolumn}
\usepackage{bm}

\begin{document}

\preprint{AIP/123-QED}

\title {Phase transition properties of Bell-Lavis model}

\author{M. \v{S}im\.enas}
\author{A. Ibenskas}
\author{E. E. Tornau}\email{et@et.pfi.lt}
\affiliation{Semiconductor Physics Institute, Center for Physical Sciences and Technology, Go\v{s}tauto 11, LT-01108 Vilnius, Lithuania.}


\date{\today}

\begin{abstract}
Using Monte Carlo calculations we analyze the order and the universality class of phase transitions into a low density (honeycomb) phase of a triangular antiferromagnetic three-state Bell-Lavis model. The results are obtained in a whole interval of chemical potential $\mu$ corresponding to the honeycomb phase. Our results demonstrate that the phase transitions might be attributed to the three-state Potts universality class for all $\mu$ values except for the edges of the honeycomb phase existence. At the honeycomb phase and the low density gas phase boundary the transitions become of the first order. At another, honeycomb-to-frustrated phase boundary, we observe the approach to the crossover from the three-state Potts to the Ising model universality class. We also obtain the Schottky anomaly in the specific heat close to this edge. We show that the intermediate planar phase, found in a very similar antiferromagnetic triangular Blume-Capel model, does not occur in the Bell-Lavis model.  

\end{abstract}

\maketitle

\section{Introduction}

Two-dimensional supramolecular structures with a planar honeycomb (HON) pattern on metal surfaces are very often encountered in molecular assemblies~\cite{bartels,barth}. Such systems are composed of triangular- or mixed triangular and rectangular-shaped organic molecules connected by hydrogen bonds. Ordered HON formations are obtained in triangular-molecule-assemblies of trimesic acid (TMA)~\cite{dmitriev,griessl,li,ye}, BTB~\cite{kampschulte,gutzler}, 1,3,5-tris(pyridine-4-ylethynyl)benzene~\cite{ciesielski}, melamine~\cite{silly,mura}, etc. Sometimes the triangular molecules are used as nodes, while PCTDI~\cite{theobald,weber,silly1}, NTCDI~\cite{perdigao}, ditopic imidic linkers~\cite{palma} or some other rectangular-shaped molecules form hexagon sides during the assembly of the HON structure. 

Theoretical studies of the stability of certain structures, the adsorption energies, the most probable ocupation sites or the interaction parameters of H-bonded molecular structures are usually performed by the density functional theory methods~\cite{mura,martsi,rochefort}. To obtain density-temperature phase diagrams or predict new ordering motifs, the statistical models of phase transitions might also be used~\cite{weber,balbas,misiunas,simenas,simenas1} as an alternative or a supporting simulation. Since the triangular molecules are symmetric and their layout is planar, they possess a three-fold symmetry with respect to $120^{\circ}$ rotation. Thus, the HON phase (Fig. 1a), organized by triangular molecules, H-bonded by their vertices (tip-to-tip bonding) on the triangular lattice, might be considered as the three-state system. If the ``leg'' of the molecule is superposed with the direction of the lattice  vector, there are just two molecular states, which differ by 60 degrees rotation, and the vacancy state. For such a definition of the states, the HON phase corresponds to a phase on a tripartite lattice with the sites of each sublattice occupied by different occupation variables (e. g. +1, $-1$, and 0).

This phase is a popular ground state structure of well-known three-state models: the Potts model~\cite{wu}, the Blume-Capel (BC) model~\cite{bc,bcafm} and the Blume-Emery-Griffiths (BEG) model~\cite{beg}. The HON phase is obtained in these models for the antiferromagnetic (AFM) nearest-neighbor interaction and the certain range of a single-ion anisotropy parameter which corresponds to a chemical potential, if the lattice-gas formalism is used (e.g. in case of molecular ordering). In a limiting range of the HON phase existence (for low values of chemical potential or high molecular concentrations), the centers of honeycombs are to be filled and the three-state model might be related to the triangular AFM Ising (TAFI) model, characterized by~\cite{wannier,baxter} the frustrated phase and finite entropy at zero temperature. 

Nevertheless, for a description of the tip-to-tip ordering of triangular (e.g. TMA~\cite{misiunas}) molecules, the mentioned three-state models are unsuitable. The AFM Potts model cannot be used, since it takes into account a non-zero contribution to the energy of interacting non-equivalent states. Obviously, the interaction of a vacancy with a molecule in any of the two non-zero molecular states ($\pm1$) has to be neglected. The AFM BC model is inappropriate, because only one of the two AFM interactions (tip-to-tip (Fig. 1b) and side-to-side (Fig. 1c)) has to be considered in a tip-to-tip ordering scheme, while the AFM BC model (as well as more general AFM BEG model) does not segregate between them accounting both as the same interaction. Moreover, the ferromagnetic (FM) interaction of two molecules in the same state, instead of giving no contribution to the energy (no H-bond), increases the energy. 

\begin{figure}[] 
\includegraphics[width=0.8\columnwidth]{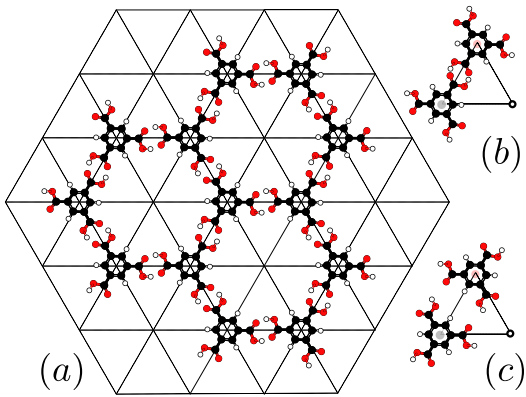} 
\caption{(color online) Schematic representation of interactions and ordering in the BL model using triangular TMA molecules as an example: (a) a fragment of the HON phase, (b) tip-to-tip double H-bond interaction ($-\epsilon_H-\epsilon_{vdW}$), (c) side-to-side interaction ($-\epsilon_{vdW}$). The C, O and H atoms of TMA molecule are shown by black, red and white circles, respectively. In (b) and (c) the sites of sublattices A, B and C are shown by gray, pink and open dots, respectively.}  
\label{fig1.eps}
\end{figure} 

There is one statistical lattice model which perfectly fits for a description of the ordering of planar triangular molecules. The model was proposed~\cite{bell} in 1970 by Bell and Lavis (BL) to consider a two-dimensional bonded fluid. This model emphasizes the orientational property of the hydrogen bond on the triangular lattice, and therefore is often considered~\cite{barbosa,fiore} as a lattice version of the Mercedez-Benz models~\cite{MB}. 
According to the BL model, the H-bonded pair of molecules is created, if the bonding directions of two nearest-neighbor (NN) molecules point towards each other. Otherwise, there is just van der Waals interaction between the NN molecules. At lower temperature, the phase transition (``long-range sublattice order'') occurs in the BL model between a disordered ``liquid'' and a ``solid'' or ``ice'' phase. From a magnetic point of view the BL model is clearly antiferromagnetic, since the H-bond occurs only when two molecules are in different states (molecular orientations). Thus, the mentioned phase transition is between the disordered paramagnetic (PM) and the HON phases. 

The thermodynamic properties of the BL model were studied in some detail using mean-field~\cite{bell}, renormalization group~\cite{young}, cluster variation~\cite{bruscolini}, recursive
approach to the Bethe lattice~\cite{barbosa} and Monte Carlo~\cite{patrykiejew,fiore,misiunas} methods. 
It was shown~\cite{young,barbosa} that the BL model can be mapped into the AFM BEG model if an anisotropic interaction is added to the BEG model Hamiltonian. Thus, though the properties and resulting phase diagrams
might have some similarities, there are some important differences between the BEG and BL models (see Ref.~\cite{barbosa} for more detail). 

At least two important problems of the BL model were not thoroughly studied in previous papers. The first deals with a recent finding~\cite{zukovic} of the Berezinskii-Kosterlitz-Thouless (BKT)-type phase between the disordered and long-range ordered (HON) phase in the AFM BC model. This frustrated phase does not cease to exist even in the AFM BC model with exclusions~\cite{ibenskas}. In absence of van der Waals interactions, the BL model has the same edges of the single ion anisotropy (chemical potential) for the HON phase existence as the AFM BC model. With an increase of molecular density, both BL and AFM BC models approach the two-state (TAFI) model limit. Moreover, the two peaks are observed in the temperature dependence of the specific heat of the BL model at higher molecular densities - the same finding as for the AFM BC model. The question then arises: how similar are the phase transitions of those two models? Do the peaks of the specific heat frame the planar phase or their origin is different?

It should be noted, that the relation between the frustrated phase of the TAFI model and the planar-type of ordering is well-known. The magnetic field (chemical potential) breaks the ground state degeneracy of the TAFI model and stimulates the occurrence of the  $\sqrt3\times\sqrt3$ structure at low values of temperature~\cite{schick}. The system splits into three sublattices with ferrimagnetic spin magnetizations $m_{\mathrm{A}}=m_{\mathrm{B}}\neq m_{\mathrm{C}}$ (A, B and C are the three sublattices of the triangular lattice). The phase transition to this phase is shown~\cite{alexander} to belong to the three-state Potts universality class. The mapping of the TAFI model in a field to the solid-on-solid model~\cite{blote1} leads to a prediction of the BKT phase transition point~\cite{KT} at $T=0$.

With added next-nearest-neighbor (NNN) FM interaction, both the three-state AFM Potts model~\cite{ono} and the TAFI model~\cite{landau,blote,takayama,miyashita} are known to have BKT-type phase transitions in the same universality class as the six-state clock model~\cite{jose,tobochnik,noh}. Under simple transformation the six-state clock model is mapped to the TAFI model~\cite{miyashita}. The six-state model can exhibit either a first-order transition, two BKT-type transitions, or successive Ising, three-state Potts, or Ashkin-Teller-type
transitions~\cite{cardy}. The competition of the NN AFM interactions and NNN FM interactions in the TAFI model leads to a two-peaked temperature dependence of a specific heat~\cite{landau} framing the planar phase in a similar manner as for the $q$-state clock models~\cite{lapili}. 
 
The second unsolved problem of the BL model is related to the order and the universality class of the phase transition from the disordered to the HON phase. The previous results are rather controversial. The first-order phase transition was initially obtained~\cite{bell} using a mean-field approximation. Later, the calculation based on the renormalization group approach for partial BL model had predicted~\cite{young} the transition being in the FM rather than the AFM three-state Potts universality class~\cite{baxter} (note, that a (weak) first order phase transition was later determined~\cite{adler} for the three-state AFM Potts model). The authors~\cite{young} of this prediction even assumed that the second-order phase transition found in their work might be a consequence of the low dimensionality of the BL model. In Refs.~\cite{bruscolini,barbosa} the BL model was solved by the cluster variation method and a weak first-order phase transition was again obtained. Later Fiore et al~\cite{fiore} performed Monte Carlo (MC) calculations and attributed the transition to the Ising universality class, claiming that the first order phase transition obtained in previous papers was the artifact arising due to the Bethe-like (cluster) methods used for calculation. 

Here we thoroughly study these two problems using MC simulation and finite size scaling. We found that despite many similarities (same ground state HON structure; same limits of the HON phase and proximity of the frustrated phase edge) the BKT-type planar phase does not occur in the BL model. The low temperature peak of the specific heat is shown to be caused by the Schottky anomaly. The obtained critical exponents clearly demonstrate that the transition to the HON phase belongs to the three-state Potts universality class.

\section{Model and details of simulation}

The BL model is based on the assumption that each molecule has three bonding directions at $120^{\circ}$ angle to each other. The molecule has three states: two orientational states and a vacancy state. In each of orientational states the molecule has bonding directions pointing towards three of the six NN sites of the triangular lattice. The H-bond is formed if two NN molecules point towards each other. The interaction energy of a pair of molecules at NN sites is $-\epsilon_H-\epsilon_{vdW}$ and $-\epsilon_{vdW}$ for H-bonded and H-unbonded pair, respectively, and the subscripts H and vdW denote H-bond and van der Waals interactions, respectively (see Fig. 1).

Several alternative representations of the BL model Hamiltonian are known. The spin-1 variables and mapping scheme to the BEG model were suggested by Young and Lavis~\cite{young}. This scheme was later employed by Barbosa et al~\cite{barbosa} who used convenient lattice-gas variables to separately describe occupational and orientational ordering (see, also~\cite{fiore}). In Ref.~\cite{misiunas} the description in terms of bond vectors and corresponding energy functionals was used. Gorbunov et al~\cite{gorbunov} employed Kronecker variables to express the Hamiltonian of the BL model. Here we use the lattice-gas Hamiltonian suggested by Barbosa et al~\cite{barbosa} in a form

\begin{equation}
{\cal H}= -\sum_{i,j}n_in_j(\epsilon_{vdW}+\epsilon_H\tau_i^{ij}\tau_j^{ji})+\mu\sum_{i}n_i, 
\end{equation}

where the occupation variable $n_i$ is 1 if the site $i$ is filled by the molecule and 0 if the site is empty. This variable is related as $n_i=s_i^2$ to the spin-1 model variable ($s_i=\pm1$ and 0) further used to characterize the order parameter. Another variable $\tau_i^{ij}$ is responsible for the orientational ordering: it corresponds to presence ($\tau_i^{ij}=1$) or absence ($\tau_i^{ij}=0$) of the H-bonding in the direction from site $i$ to site $j$. Here $\mu$ stands for the chemical potential which we write here with a plus sign as in the BEG model. The first sum is performed over all pairs of nearest neighbors $i$ and $j$.

Further, the temperature, chemical potential, and van der Waals interaction are normalized
to H-bond interaction, $\epsilon_H$: $k_BT/\epsilon_H$, $\mu/\epsilon_H$ and $\xi=\epsilon_{vdW}/\epsilon_H$. 

The MC simulation was performed using Metropolis
algorithm and Glauber dynamics, i.e. the calculations were carried out with fixed chemical potential $\mu$ and varying molecular coverage. Single-flip algorithm was chosen, because of a poor performance~\cite{coddington} of traditional cluster algorithms when applied to frustrated systems.
To check, if the simple single-flip technique is appropriate to study such partly frustrated systems as the BL model at $\mu\rightarrow0$, we also 
performed some calculations using Wang-Landau sampling~\cite{wang} which is considered more reliable to address the frustrated systems. The test calculations of specific heat temperature dependence at $\mu/\epsilon_H=0.1$ demonstrated a perfect agreement between the single-flip and Wang-Landau methods.  

We used the triangular lattices with periodic boundary conditions of sizes $L\times L$ with $L=24, 48, 72, 96, 120$ and 144 for thermal averaging MC calculations and finite size scaling. We discarded $10^5-10^6$ MC steps (MCS) for thermalization and collected averages of $(3-4)\times10^6$ MCS (for the edges of the HON phase existence we used to take up to $1.4\times10^7$ MCS). Our simulations were performed starting from higher temperature in the PM phase and using random initial particle configuration. Then the temperature was gradually decreased in small steps with simulations at a new temperature starting from the final configuration of the previous temperature. 

We estimated the thermalization period by observing the time evolution of the order parameters and
energy at different temperatures. Before gathering statistics for thermal averaging, we also
made additional checks at multiple temperatures near phase transition points to be sure
that the sample is in the equilibrium. The thermalization time did not exceed $2\times10^5$ MCS 
for lattice size $L=144$.

In order to estimate statistical errors, we used the data from $n\approx 5$ independent simulation runs starting from different initial states. The observation $x_i$ of each run was used to obtain a mean value $\langle x\rangle$ at that particular temperature. The error bar of $\langle x\rangle$ can be obtained from $\sigma=s/\sqrt{n-1}$
and $s^2=\frac{1}{n}\sum_{i=1}^{n}(x_i-\langle x\rangle)^2$. In Figs the errors roughly correspond to the size of a data point symbol.

To find the phase transition order and detail the thermal averaging results, we also performed energy and magnetization histogram calculations using single-histogram reweighting technique~\cite{reweight}. For these calculations we used larger lattice sizes (up to $L=192$) than for the thermal averaging and collected entries from $2\times10^7$ MCS for each histogram. In our simulations of thermodynamic parameters the phase transition point was first located by the thermal averaging and then recalculated by the histogram method. The results were considered reliable if the data obtained by both methods matched. 

We also performed the analysis of the autocorrelation time of energy at $T_c$ and some values of $\mu$. The integrated autocorrelation time ranges from $\tau\sim10^3-10^4$ MCS at $0<\mu/\epsilon_H<1.4$ ($L=144$) to
$\tau\sim10^6$ MCS at the first-order phase transitions close to the gas phase, $\mu/\epsilon_H=1.48$ ($L=72$). 

We used two AFM order parameters to study the PM-to-HON phase transition properties. One of them was the staggered magnetization (average difference of sublattice occupancy by non-zero variables) suggested for studies of the HON phase in the AFM BC model~\cite{zukovic}

\begin{eqnarray}
m_s &=& \langle M_s\rangle/L^2 =
\frac{3}{2L^2}\Big\langle\mathrm{max}\Big(\sum_{i\in\mathrm{A}}s_{i}, \sum_{j\in\mathrm{B}}s_{j},\sum_{k\in\mathrm{C}}s_{k}\Big)\nonumber\\ &-& \mathrm{min}\Big(\sum_{i\in\mathrm{A}}s_{i}, \sum_{j\in\mathrm{B}}s_{j},\sum_{k\in\mathrm{C}}s_{k}\Big)\Big\rangle.
\end{eqnarray}

Here A, B and C correspond to three sublattices of the triangular lattice. For some calculations we also used another order parameter which was simply the difference
\begin{equation}
m_{10}=3\langle M_{10}\rangle/L^2=\langle\rho_{\pm 1}-\rho_{0}\rangle,
\end{equation}
where $\rho_{\pm1}$ means the sublattice occupancy by dominating non-zero (either +1 or $-1$) variable and $\rho_{0}$ - the sublattice occupancy by dominating zero variable.

We calculated temperature dependences of the specific heat $C_v=(\langle{\cal H}^2\rangle-\langle{\cal H}\rangle^2)/L^2k_BT^2$, susceptibility $\chi_x=(\langle M_x^2\rangle-\langle M_x\rangle^2)/L^2k_BT$,  logarithmic derivatives of $\langle M_s\rangle$ and $\langle M_s^2\rangle$ 

\begin{eqnarray}
D_{1s}=\frac{\partial\ln\langle M_s\rangle}{\partial\beta}=
\frac{\langle M_s{\cal H}\rangle}{\langle\ M_s\rangle}-\langle {\cal H}\rangle,\\\nonumber
D_{2s}=\frac{\partial\ln\langle M_s^2\rangle}{\partial\beta}=\frac{\langle M_s^2{\cal H}\rangle}{\langle M_s^2\rangle} -\langle {\cal H}\rangle,          
\end{eqnarray}

and Binder order parameter and energy cumulants, $U_B^{m}=1-\langle M_s^4\rangle/3\langle M_s^2\rangle^2$ and $U_B^E=1-\langle {\cal H}^4\rangle/3\langle {\cal H}^2\rangle^2$, respectively. Here the subscript $x$ corresponds to $s$ and 10. 

Close to the second order phase transition point $T_c$ and for sufficiently large $L$, the order parameter, susceptibility and specific heat can be expressed by the scaling functions $X$, $Y$ and $Z$ in a following way 

\begin{eqnarray}
m_x&=&L^{-\beta/\nu}X(tL^{1/\nu}),\\\nonumber
\chi_x&=&L^{\gamma/\nu}Y(tL^{1/\nu}),\\\nonumber
C_v&-&C_{v0}=L^{\alpha/\nu}Z(tL^{1/\nu}),
\end{eqnarray}

where $\alpha$, $\beta$, $\gamma$ and $\nu$ are the critical exponents of specific heat, magnetization, susceptibility and correlation length, respectively, and $C_{v0}$ is the background contribution to the specific heat. The maxima of $C_v$ and $\chi_x$ at $T_c$ should scale as  $\sim L^{\alpha/\nu}$ and $\sim L^{\gamma/\nu}$, respectively. Further, in order to obtain 
$\alpha/\nu$ and $\gamma/\nu$ values, we combine the scaling of the functions (5) close to $T_c$ with calculation of these critical exponents by scaling the maximum of $C_v$ and $\chi_x$ at $T_c$. The extrema of functions $D_{1s}$ and $D_{2s}$ scale as $\sim L^{1/\nu}$~\cite{ferrenberg}. 

At the first order phase transition point, the extrema of all these functions
scale as $\sim L^{d}$ ~\cite{challa}, where $d$ is the dimensionality of the system. 

In case of the BKT-type phase transitions, the correlation length diverges as $\xi=\xi_0\exp\{a[(T_{\mathrm{BKT}}-T)/T_{\mathrm{BKT}}]^{-1/2}\}$ 
and the spin-correlation function  decays as $\langle s_is_j\rangle\sim r_{ij}^{-\eta}$, where $\eta$ is the critical exponent of the correlation function~\cite{KT}. The order parameter at the BKT-type phase transition point scales as $m_x(L)\sim L^{-\eta/2}$. 

\begin{figure}[] 
\includegraphics[width=1.0\columnwidth]{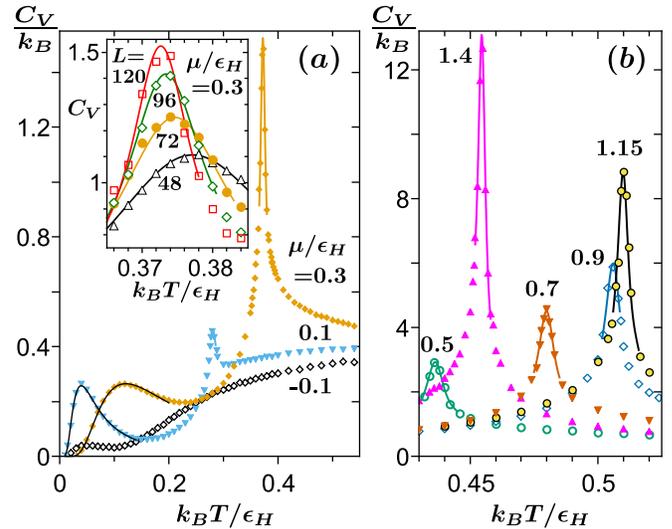} 
\caption{(color online) Temperature dependence of the specific heat at $\xi=0$ and $L=120$ and different values of chemical potential: (a) $\mu/\epsilon_H=-0.1, 0.1$ and 0.3 and (b) 0.5, 0.7, 0.9, 1.15 and 1.4. Inset in (a): $C_v(T)$ dependence at $\mu/\epsilon_H=0.3$ for different values of $L$. The points are obtained by thermal averaging and the curves are obtained by reweighted energy histogram method. The black curves used to fit the low temperature peaks at $\mu/\epsilon_H=0.1$ and 0.3 in (a) are the results of the 2-level model (see text).}  
\label{fig2.eps}
\end{figure} 

\begin{figure}[] 
\includegraphics[width=1.0\columnwidth]{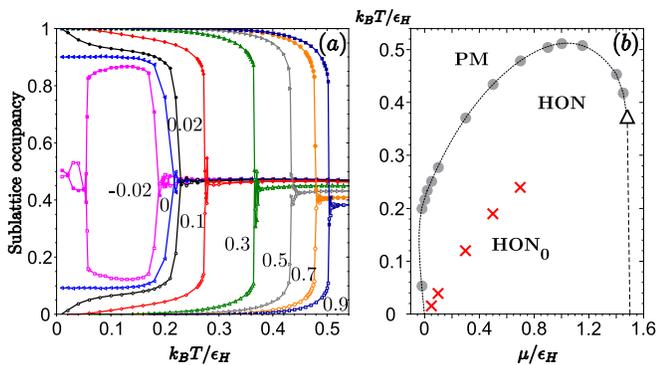} 
\caption{(color online) The BL model at $\xi=0$ and $L=120$: (a) Occupancy of the A (B) sublattice by +1 (-1) states (higher branch) and -1 (+1) states (lower branch) for $\mu/\epsilon_H=0.9$, 0.7, 0.5, 0.3, 0.1, 0.02, 0 and -0.02. (b) Phase diagram obtained from temperature dependence of the specific heat. The solid circles are $T_c$ points, red crosses - points of Schottky anomaly, triangle - tricritical point. The denotation HON corresponds to sublattice occupancies $\rho_{\mathrm{A}}=\rho_{\mathrm{B}}$, $\rho_{\mathrm{C}}>0$ and HON$_0$ (``perfect'' HON phase) corresponds to 
$\rho_{\mathrm{A}}=\rho_{\mathrm{B}}\approx1$, $\rho_{\mathrm{C}}\approx0$.}
\label{fig3.eps}
\end{figure} 

\begin{figure}[] 
\includegraphics[width=1.0\columnwidth]{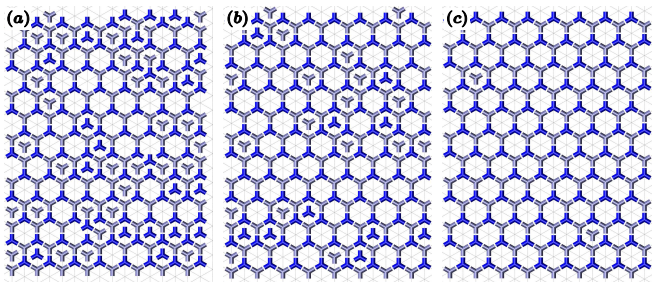} 
\caption{(color online) Snapshots of the HON structure fragment at $\mu/\epsilon_H=0.3$ and below $T_c$ point: (a) $k_BT/\epsilon_H= 0.35$, (b) 0.2, (c) 0.06. The sublattices A and B are almost occupied by +1 (blue tripod-shaped symbol) and -1 (gray tripod-shaped symbol) states, respectively, and the system undergoes gradual emptying of the sublattice C with decrease of temperature.}  
\label{fig4.eps}
\end{figure}

\section{Results}

Ground state calculations demonstrate that the ordered low density structure (HON phase) is obtained between the values of chemical potential $\mu/\epsilon_H=6\xi$ (frustrated phase-HON phase boundary) and $\frac{3}{2}(1+\xi)$ (HON-gas phase boundary). If the van der Waals interactions are neglected ($\xi=0$), the limits of the HON phase are between 0 and 3/2, and this range coincides with the range the HON phase occupies in a similar triangular AFM BC model~\cite{bcafm}. 
The AFM BC model demonstrates~\cite{zukovic} two consecutive phase transitions: from the disordered (PM) to the BKT-type phase and from the BKT-type phase to the HON phase. 

In Figs. 2a and 2b we present temperature dependences of the specific heat of the BL model for $\xi=0$ and different values of chemical potential $\mu$. 
At higher values of $\mu$, one peak of specific heat related to the disordered-to-HON phase transition at $T_c$ is observed. The magnitude of the peak decreases with decrease of $\mu$. The second peak at low temperature starts to emerge for $\mu/\epsilon_H\leq 0.5$ (see $C_v(T)$ at $\mu/\epsilon_H=0.3$ and 0.1 in Fig. 2a). This low-temperature anomaly exists at some higher $\mu$ values as well, but it cannot be seen due to its relative smallness and proximity to the peak at $T_c$. The temperature dependence of high temperature $C_v$ peak depends on $L$ as shown in inset of Fig. 2a, but the low temperature anomaly is clearly $L$-independent.

\begin{figure}[] 
\includegraphics[width=1.0\columnwidth]{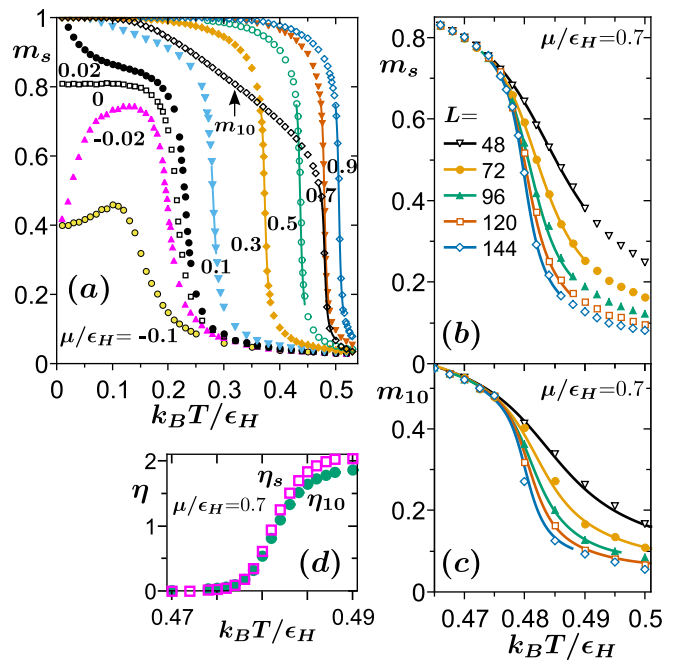} 
\caption{(color online) Temperature dependences of order parameters at $\xi=0$ and $L=120$: (a) $m_s$ at $\mu/\epsilon_H=-0.1 - 0.9$ and $m_{10}$ at $\mu/\epsilon_H=0.7$, (b) $m_s$ and (c) $m_{10}$ at $\mu/\epsilon_H=0.7$ and different values of $L$, and (d) temperature dependences of exponents $\eta_s$ and $\eta_{10}$ obtained at $\mu/\epsilon_H=0.7$ using order parameters $m_s$ and $m_{10}$, respectively. In (a-c) the points are the results of thermal averaging and curves are obtained from reweighted magnetization histograms.}  
\label{fig5.eps}
\end{figure} 

\begin{figure}[] 
\includegraphics[width=1.0\columnwidth]{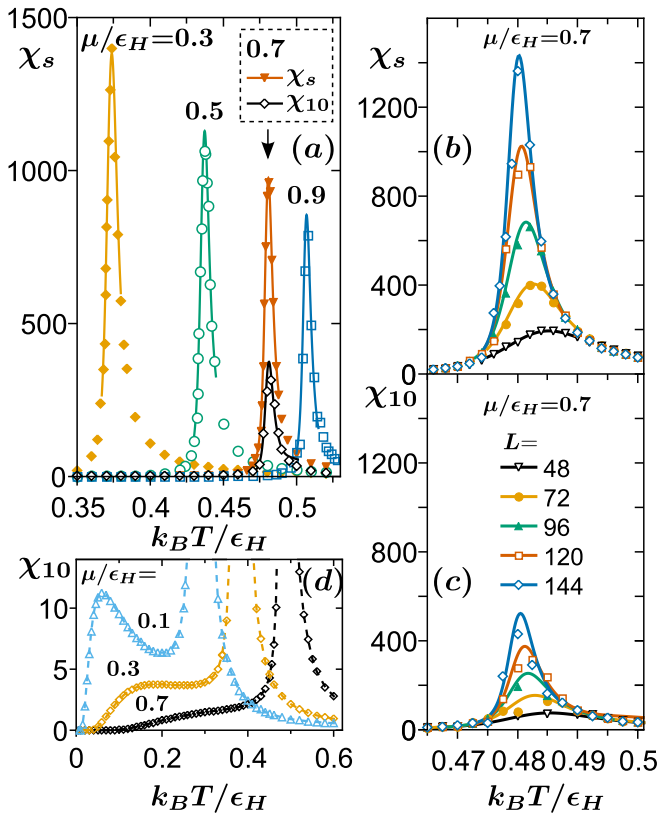} 
\caption{(color online) Temperature dependences of susceptibility at $\xi=0$ and $L=120$: (a) $\chi_s$ at $\mu/\epsilon_H=0.9, 0.7, 0.5$ and 0.3 and $\chi_{10}$ at $\mu/\epsilon_H=0.7$ close to $T_c$, (b) $\chi_s(T)$ and (c) $\chi_{10}(T)$ at $\mu/\epsilon_H=0.7$ and different values of $L$, (d) $\chi_{10}$ at $\mu/\epsilon_H=0.7, 0.3$ and 0.1 close to Schottky anomaly point $T_S$ ($\chi_{10}(T_S)\ll\chi_{10}(T_c)$). The points are the results of thermal averaging and the curves are obtained by reweighted magnetization histograms.}  
\label{fig6.eps}
\end{figure}

The low-temperature peak is due to the Schottky anomaly. It appears at $\mu\rightarrow0$, because of the scarcity of spin configurations and the proximity of the ground state and exited energy levels. The situation is analogous to that seen in the 1D Ising model at finite temperature. The occurrence of the Schottky peak in the BL model at low $T$ might be described by a simple 2-level energy model. Below the phase transition point $T_c$, the two sublattices become occupied by the molecules in +1 and $-1$ states, respectively. The third sublattice is partly occupied by molecules at $T<T_c$, but it gradually empties with a decrease of temperature (see Fig. 3a and series of snapshots in Fig. 4). The coverage of this sublattice depends on $\mu$ and is higher at small $\mu/\epsilon_H\lesssim0.5$ (note, that at large $\mu$ the coverage of the third sublattice is low and it empties almost immediately at $T_c$). The peak of the Schottky anomaly coincides with the absence of molecules in a third sublattice and establishment of a ``perfect'' HON phase (HON$_0$ in the phase diagram of Fig 3b). Thus, above this peak there are roughly two levels of local energy only: 0 (if the center of the hexagon, formed of alternating $\pm1$ variables in other two sublattices, is empty) and $\mu/\epsilon_H$ (if the center of the hexagon is filled). Note, that the energy does not depend on a state of a non-zero variable in the hexagon center. The specific heat of the 2-level system is equal to $C_v/k_B=\delta^2\exp(-\delta)/[1+\exp(-\delta)]^2$, where $\delta=\Delta/k_BT$ and $\Delta$ corresponds to the difference between two energy levels.
We tried to fit the $C_v(T)$ dependence of the BL model in the vicinity of the low temperature peak by the 2-level model and obtained a very reasonable agreement at low values of $\mu$ (see Fig. 2a). Our curves for $\mu/\epsilon_H=0.1$ and 0.3 are described by the 2-level model formula with $\Delta/\epsilon_H=0.095$ and 0.29, respectively (to fit the peak heights we multiplied the specific heat of the 2-level model by 0.6).

In Figs. 5a and 6a the temperature dependences of order parameters $m_s$ (2) and $m_{10}$ (3) and their corresponding susceptibilities are presented at different values of $\mu$. It is seen that $m_s(T)$ dependence steepens and the peak of the susceptibility $\chi_s(T_c)$ becomes sharper and decreases in height with increase of $\mu$. In Fig. 6a we limit ourselves by the value of chemical potential $\mu/\epsilon_H\leq0.9$. With further increase of $\mu$, the peak height decreases up to $\mu_c/\epsilon_H\approx1.1$ and then starts to increase again. Here $\mu_c$ marks the value of chemical potential at the top of the $(T,\mu)$ phase diagram (Fig. 3b), and $T_c$ starts to decrease at $\mu>\mu_c$. The peaks of $\chi_s(T_c)$ and $C_v(T_c)$ are very sharp and high at $\mu>\mu_c$ (see the peak at $\mu/\epsilon_H=1.4$ in Fig. 2b). The reason is that the phase transitions in the interval of chemical potential above $\mu>\mu_c$  are close to the tricritical region. The $L$-dependences of $m_s$ and $\chi_s$ close to $T_c$ are presented in Figs. 5b and 6b, respectively. They are further used for the finite size analysis.

In order to check if the BKT-type phase exists between two anomalies of the specific heat, we performed the calculation of the critical exponent of the correlation function, $\eta$. 
To obtain $\eta_s$, the slopes of $\ln m_s(L)$ vs $\ln L$ at different temperature around the $T_c$ point were calculated. It should be noted, that the order parameter identical to $m_s$ (2) was used to obtain the BKT-type critical line for the AFM BC model~\cite{zukovic,ibenskas}. As is seen from $\eta_s(T)$ dependence at $\mu/\epsilon_H=0.7$ (Fig. 5d), there is no plateau which would indicate the existence of the critical line of the BKT-type points in the BL model. We calculated $\eta$ for $\mu/\epsilon_H=0.3$ as well, but the result was the same. In comparison, in the AFM BC model the plateau of $\eta(T)$ dependence, corresponding to the BKT-type transitions, was found~\cite{zukovic} for $\eta$ in between 0.12 and 0.29. The limits of the planar phase for a classical planar rotator model with sixfold symmetry breaking fields~\cite{jose} were in between 1/9 and 1/4.

\begin{figure*}[] 
\includegraphics[width=1.7\columnwidth]{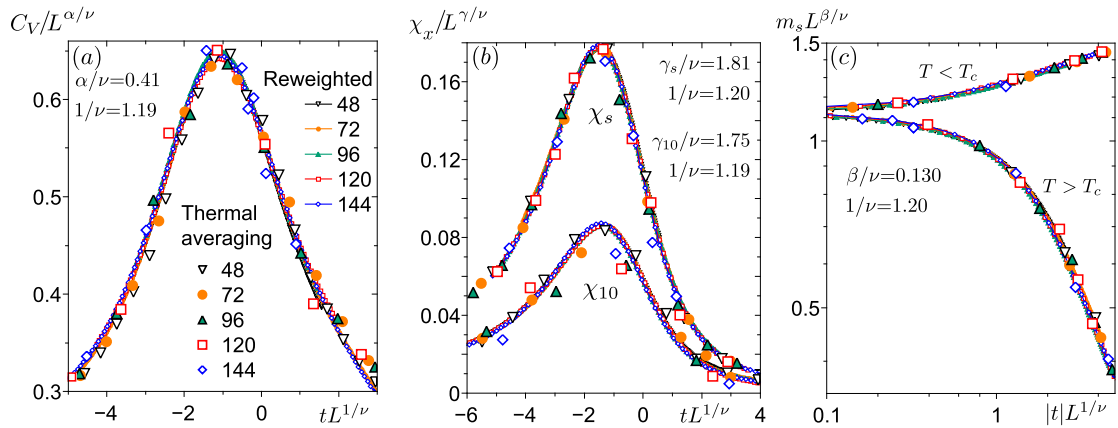} 
\caption{(color online) Finite-size scaling of (a) specific heat, (b) both susceptibilities and (c) order parameter $m_s$ at $\mu/\epsilon_H=0.7$ The results are fitted using formulas (5), $t=(T_c-T)/T_c$ and the background of $C_v$ is assumed to be zero. Large symbols correspond to the results of thermal averaging; lines correspond to the results obtained close to $T_c$ by the reweighted histogram method.}  
\label{fig7.eps}
\end{figure*}

\begin{table}
\caption{The critical exponents of the BL model.}
\begin{ruledtabular}

\begin{tabular}{llllllllll}
$\xi$ & $\mu/\epsilon_H$ & $\alpha$ & $\gamma$ & $\beta$ & $1/\nu_{[\mathrm{D_{1s}}]}$ & $1/\nu_{[\mathrm{D_{2s}}]}$ & $1/\nu_{[C_v]}$\\
\hline
  &        & $\pm0.01$ &$\pm0.02$&$\pm0.005$ & $\pm0.02$& $\pm0.02$& $\pm0.02$     \\
0 & 0.05   & 0.11   &   &         & 0.99 & 1.06 & 1.00\\
  & 0.10   & 0.16   & 1.67  &   0.119 & 1.10 & 1.09&  1.11\\
  & 0.30   & 0.29   & 1.58  &   0.112 & 1.20 & 1.20&  1.20\\
  & 0.50   & 0.33   & 1.53  &   0.112 & 1.20 & 1.21& 1.21\\
  & 0.70   & 0.34   & 1.51  &   0.108 & 1.22 & 1.22& 1.19 \\
  & 0.70$^a$       &        & 1.47  &   0.11  & 1.18 & 1.19&      \\
  & 0.90   & 0.35   & 1.47  &   0.111 & 1.18 & 1.18& 1.20 \\
  & 1.15   & 0.40   & 1.47  &   0.109 & 1.24 & 1.24& 1.20\\
  & 1.40   & 0.47   & 1.34  &   0.108 & 1.22 & 1.23& 1.20\\
  & 1.45   & 0.57   & 1.15  &         & 1.38 & 1.41& 1.35   \\
  & 1.48$^b$   & 1.79$\nu$ & 2.12$\nu$ & & 2.11 & 2.18 & 2.00\\  

\hline
0.1 & 0.70 & 0.18  &  1.64  &  0.116  & 1.11 & 1.09 &\\
    & 0.90 & 0.30  &  1.51  &  0.106  & 1.22 & 1.21 &\\
    & 1.10 & 0.33  &  1.48  &  0.108  & 1.20 & 1.20 &\\
    & 1.30 & 0.35  &  1.47  &  0.106  & 1.22 & 1.23 &\\
    & 1.40 & 0.35  &  1.49  &  0.106  & 1.18 & 1.18 &\\
\hline
Ising &  & 0    & 7/4 & 1/8 & 1 &  & &\\
Potts &  & 1/3 & 13/9 & 1/9 & 6/5 &  & &

\end{tabular}
\end{ruledtabular}
The exponents $\alpha$, $\gamma$ and $\beta$ were obtained by scaling (5) of the functions $C_v$, $\chi_s$, $m_s$, respectively, close to $T_c$. The $1/\nu$ [$C_v$] is obtained from scaling of the $C_v(T)$. To adjust the values of $\alpha$ and $\gamma$, we performed the scaling of the maxima of $C_v$ and $\chi_s$, respectively. The $1/\nu [D_{1s}]$ and $1/\nu [D_{2s}]$ were obtained by scaling of the extrema of functions $D_{1s}$ and $D_{2s}$.\\$^a$ The scaling of $\chi_{10}$, $D_{1s}$ and $D_{2s}$ was performed using order parameter $m_{10}$. 
$^b$ The exponents have no sense for the first order phase transition, here we present their ratios.
\end{table}

The order parameter $m_s$ (2) does not show the Schottky anomaly at low temperature, since $m_s$ is constructed of non-zero variables. At the same time, the order parameter $m_{10}$ (3), which accounts for the emptying of the ``zero`` sublattice, is quite sensitive to the low temperature anomaly at small values of $\mu$. Its susceptibility, $\chi_{10}$, demonstrates (Fig. 6d) either a small peak (at $\mu/\epsilon_H<0.5$) or some kind of ''shoulder`` (at higher $\mu/\epsilon_H$), both of them being around two orders of magnitude smaller than the main peak at $T_c$ (cf Figs. 6a and d). Both $m_{10}(T)$ and  $\chi_{10}(T)$ do not depend on $L$ in the vicinity of the Schottky anomaly, contrary to their behavior close to $T_c$ (Figs. 5c and 6c).

Consider the $m_{10}(T)$ dependence at $\mu/\epsilon_H=0.7$ (Fig. 5a) in more detail. Such a behavior of $m_{10}(T)$ contributes to two anomalies of its susceptibility: the peak at $T_c$ (the same point as given by the order parameter $m_s$, see Fig. 6a) and the small shoulder at the Schottky anomaly point, $k_BT/\epsilon_H\leq 0.2$ shown in Fig. 6d. Roughly, at this temperature point the $m_{10}(T)$ curve saturates in Fig. 5a indicating perfect emptying of the C sublattice. 

\begin{figure}[] 
\includegraphics[width=1.0\columnwidth]{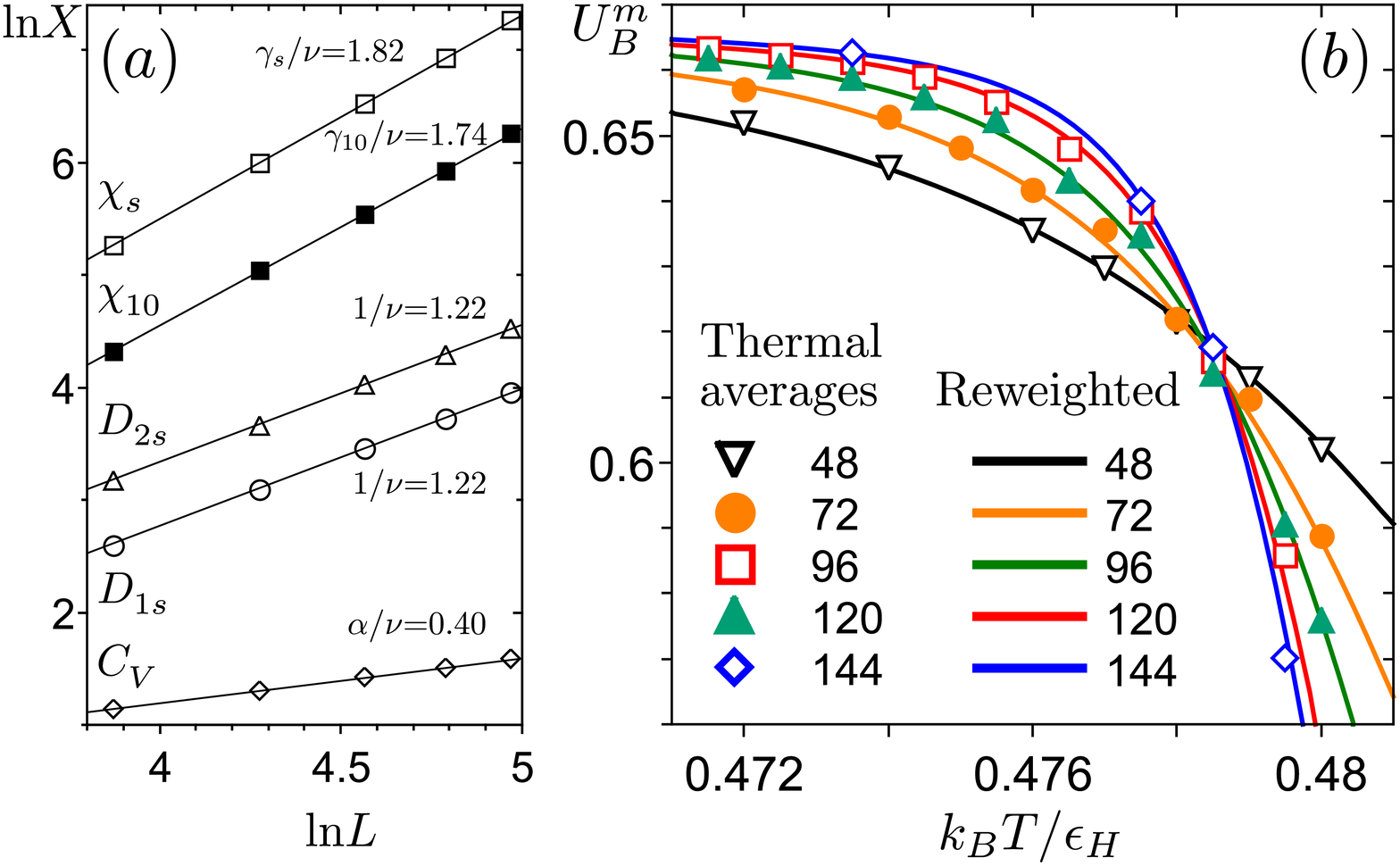} 
\caption{(color online) (a) The log-log dependences of $X=\chi_s, \chi_{10}, D_{1s}$, $D_{2s}, C_v$ vs $L$ at $T_c$; (b) temperature dependence of the Binder cumulant of magnetization $m_s$ at $T_c$. The parameters: $\mu/\epsilon_H=0.7$ and $L=120$.}  
\label{fig8.eps}
\end{figure} 

We performed the calculation of the critical exponent of the correlation function, $\eta_{10}$, using temperature and $L$ dependence of the order parameter $m_{10}$ (Fig. 5c). And again, as for the order parameter $m_s(T)$, no plateau of the BKT-type transitions was found in the $\eta_{10}(T)$ dependence (Fig. 5d).

Further, to determine the universality class of the disordered-to-HON transition, using formulas (5) we performed the finite size scaling of thermodynamic parameters at $T_c$. The scaled curves of specific heat, susceptibility $\chi_s$ and magnetization $m_s$ for $\mu/\epsilon_H=0.7$ are shown in Fig. 7a, b and c, respectively. The data used was obtained from both thermal averaging and histograms reweighting. The values of critical exponents of model (1) for different values of $\mu$ are presented in a Table I. We have found that the transition has to be attributed to the three-state FM Potts universality class in a whole range of $\mu$ values, except for the boundaries of the HON phase. 

\begin{figure*}[] 
\includegraphics*[width=2.0\columnwidth]{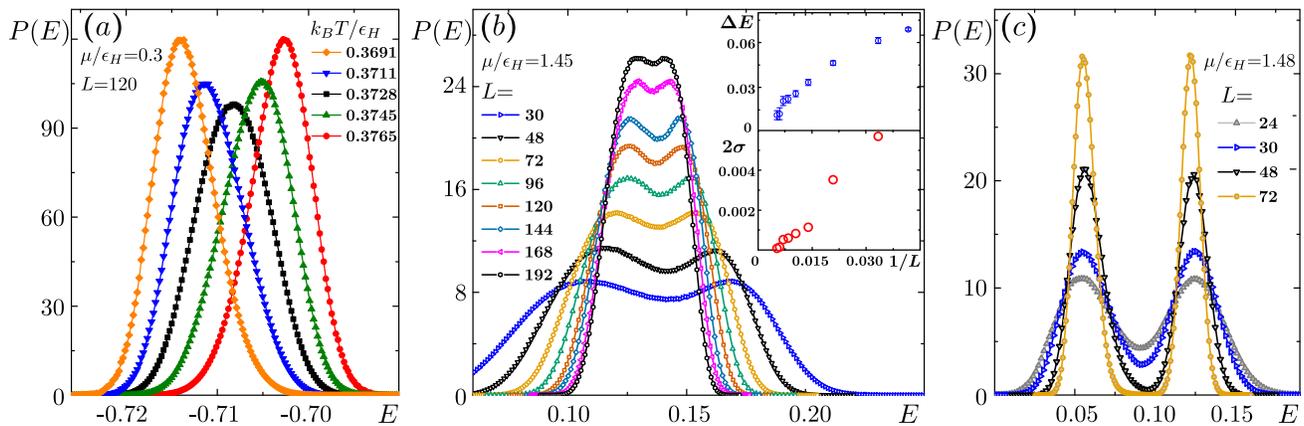}
\caption{(color online) Energy histograms: (a) at $\mu/\epsilon_H=0.3$ and $L=120$ close to phase transition ($k_BT_c/\epsilon_H=0.3728$) temperature, (b) at $\mu/\epsilon_H=1.45$ and (c) 1.48 for different lattice sizes at $T_c$. Insets in (b): lattice size dependence of interface tension and latent heat at $\mu/\epsilon_H=1.45$.}  
\label{fig9.eps}
\end{figure*} 

Our conclusion is mostly based on values of $\alpha/\nu$ and $1/\nu$, because the difference between the 3-state Potts ($\alpha=1/3$, $\nu=5/6$, $\beta=1/9$ and $\gamma=13/9$~\cite{baxter}) and Ising ($\alpha=0$, $\nu=1$, $\beta=1/8$ and $\gamma=7/4$) values for other ratios of critical exponents is of the order of the calculation error (cf $\gamma/\nu=1.733$ and 1.75 in Potts and Ising universality class, respectively). 
The scaling of the $C_v$ maximum gives the value of the exponents ratio $\alpha/\nu$ either equal (see Fig. 8a) or rather close to the Potts value, 0.4. The finite size scaling of $C_v$ using formula (5) gives the exponent $\alpha$ close to 1/3 in the range $0.3\lesssim\mu/\epsilon_H\lesssim1$ (see Table I), i.e. almost up to the top of phase diagram in Fig. 3b. No tendency towards the Ising value, $\alpha\rightarrow 0$, is noticed in this range. The scaling of the $\chi_s$ maximum at $T_c$ gives a nice linear $\ln\chi_s$ vs $\ln L$ dependences (Fig. 8a): the susceptibility exponent $\gamma_s$, obtained using order parameter $m_s$, shows some deviation from Ising-Potts values, especially close to the HON-frustrated phase boundary (see also Table I); the ratio  $\gamma_{10}/\nu=1.74$, obtained from the order parameter $m_{10}$ at $\mu/\epsilon_H=0.7$, does not allow to differentiate between the Ising and Potts values. However, the scaling of both $\chi_s$ and $\chi_{10}$ (Fig. 7b) clearly shows that the exponent $1/\nu$ has to be close to the Potts value, 1.2. This result is corroborated by the scaling of the specific heat (Fig. 7a and Fig. 8a), magnetization $m_s$ (Fig. 7c) and parameters $D_{1s}$ and $D_{2s}$ (Table I and Fig. 8a). All these parameters steadily scale with the value of $1/\nu\approx1.2$ in the range $0.3<\mu/\epsilon_H<1.4$.

The scaling of the order parameter $m_s$ is not accurate enough to unambiguously discriminate between the values of $\beta$ characteristic to the Potts and Ising universality class. Nevertheless, the scaling with the Potts model pair of ratios ($1/\nu=1.2$, $\beta/\nu=0.133$) is much better than that with the corresponding Ising model pair ($1/\nu=1$, $\beta/\nu=0.125$) for all values of $\mu$. For $\mu/\epsilon_H=0.7$ the exponent ratios are equal to $1/\nu=1.2$, $\beta/\nu=0.130$ (Fig. 7c). A slight deviation of our results from the generic relation for critical exponents ($2-\alpha=2\beta+\gamma$) is mostly due to a systematic deviation of exponent $\gamma_s$. 

In Fig. 8b we present Binder magnetization cumulant, $U_B^m$ at $T_c$. The crossing of this cumulant at $U_B^m\approx 0.61$ was considered~\cite{fiore} as the indication of the Ising behavior. However, the crossing of $U_B^m$ in the three-state Potts model also occurs at approximately the same value (see e.g.~\cite{tome}). We obtained the crossing at $U_B^m=0.615\pm0.005$.

The values of the critical exponents start to vary close to both edges of the HON phase. While
approaching the disordered-to-HON edge ($\mu/\epsilon_H\rightarrow 3/2$), we observe an increase of the exponent ratios $\alpha/\nu$, $1/\nu$ and $\gamma/\nu$ towards the value close to 2 (see Table I at $\mu/\epsilon_H=1.48$). This clearly indicates the first order phase transition. In order to determine the behavior at the tricritical region, we calculated the energy histograms presented in Fig. 9. The histogram at $\mu/\epsilon_H=0.3$ (Fig. 9a) shows a peak which moves with $T$ along the phase transition region. Such a behavior of the histograms is characteristic for almost all range of chosen $\mu$ values. It evidences a typical second order phase transition. However, close to the disordered-to-HON edge ($\mu/\epsilon_H>1.4$), the histograms are two-peaked with a high saddle point. At $\mu/\epsilon_H=1.45$ (Fig. 9b) the two-peaked histogram transits into the one peaked histogram with an increase of $L$: the latent heat $\Delta E$ and interface tension $2\sigma$ approach zero, and the transition is still of the second order. Here $\Delta E=|E_{+}-E_{-}|$, where  $E_{+}(L)$ and $E_{-}(L)$ are the energies at right and left peaks of the energy distribution, respectively, and $2\sigma=\ln[P_{\mathrm{max}}(L)/P_{\mathrm{min}}(L)]/L$, where $P_{\mathrm{max}}(L)$ and $P_{\mathrm{min}}(L)$ are the probability density of energy at the maximum and saddle point, respectively. However, at  $\mu/\epsilon_H=1.48$ (Fig. 9c) the histograms clearly indicate that the transition is of the first-order. We believe that the tricritical point is around $\mu/\epsilon_H=1.47$. The tricritical points of this model were already found at $\xi=0.1$ and 0.25~\cite{barbosa,fiore}. The location of the tricritical point and  behavior of the BL model at $\xi=0$ at the disordered-to-HON boundary, in general, is analogous to that of the AFM BC model~\cite{zukovic,ibenskas} at the same boundary.

An interesting situation is observed at another edge. At $\xi=0$, the ground state boundary between the HON and frustrated phase is at $\mu=0$. This value also marks the termination of the three-state model (1), because the third (vacant or partly vacant) sublattice becomes completely occupied at $\mu<0$. The phase existing at $\mu\leq0$ is only partly frustrated, since it retains some preference of $+1 (-1)$ states at sublattices A (B).

As shown in Table I and Fig. 10, the critical exponents $\alpha$ and $1/\nu$ (the latter obtained both from fitting of specific heat and parameters $D_{1s}$, $D_{2s}$) demonstrate systematic approach to the Ising model values starting from  $\mu/\epsilon_H=0.3$ and all the way down towards $\mu=0$. The $\alpha$  decreases from 1/3 at $\mu/\epsilon_H\geq0.5$ to 0.29 ($\mu/\epsilon_H=0.3$), 0.16 (0.1) and 0.11 (0.05), and the $1/\nu$ decreases from 1.2 at $\mu/\epsilon_H\geq0.5$ to 1 at lower values of $\mu$. However, the peaks at $T_c$ are too small and very large lattice sizes are needed to perform a reliable scaling analysis for $\mu/\epsilon_H<0.05$. The critical index $\gamma$ is less reliable at this edge, since the susceptibility fits rather badly starting from $\mu/\epsilon_H\leq0.05$. Nevertheless, our results for $\alpha$ and $1/\nu$ obtained at $0.05<\mu/\epsilon_H<0.3$ (see Table and Fig. 10) imply that close to the HON-to-frustrated phase boundary the phase transition demonstrates the approach to crossover from the three-state Potts to the Ising universality class.

\begin{figure}[] 
\includegraphics[width=1.0\columnwidth]{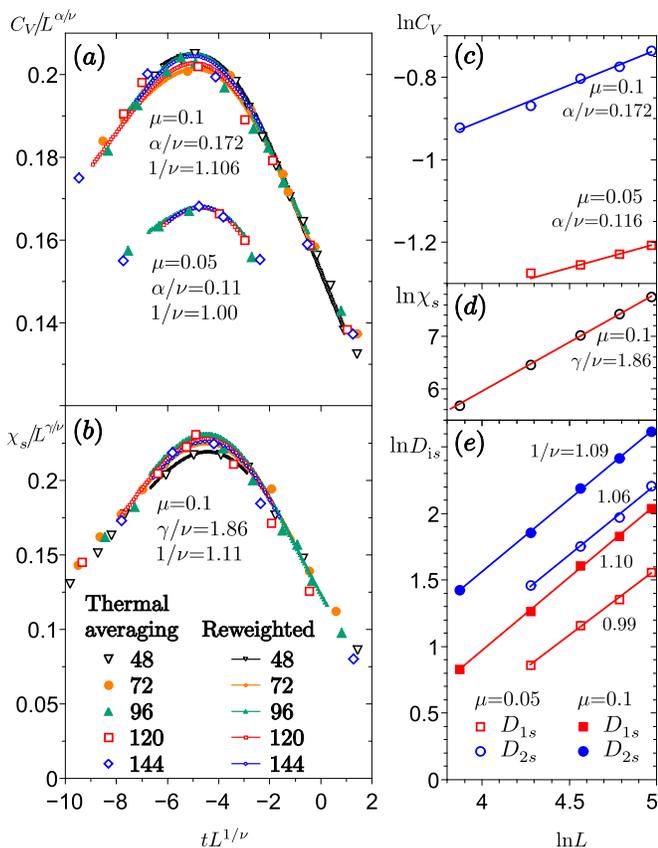} 
\caption{(color online) Finite-size scaling of (a) specific heat at $\mu/\epsilon_H=0.1$ and 0.05, (b) susceptibility $\chi_s$ at $\mu/\epsilon_H=0.1$. The log-log dependences of $C_v$ (c), $\chi_s$ (d) and $D_{1s}$, $D_{2s}$ (e) vs $L$ at $T_c$. The results are fitted using formulas (5), $t=(T_c-T)/T_c$ and the background of $C_v$ is assumed to be zero. Large symbols correspond to the results of thermal averaging; lines correspond to the results obtained close to $T_c$ by the reweighted histogram method.}  
\label{fig10.eps}
\end{figure} 

Note also, that the $T_c$ point exists for some very small $\mu<0$ values. The HON phase is still intact at finite temperature, though the ground state belongs to the frustrated phase (see Figs. 3a, b for $\xi=0$ and $\mu/\epsilon_H=-0.02$; also the phase diagrams for $\xi=0.1$ and 0.25 in Refs.~\cite{barbosa,fiore}). The existence of the HON phase at very low negative $\mu<0$ is neither spurious, nor finite size effect. It is seen comparing the $m_s(T)$ function at $\mu/\epsilon_H=-0.02$ (reentrance) and $\mu/\epsilon_H=-0.1$ (frustrated phase) at low temperature and different values of $L$. At the bump of $m_s(T)$ dependence (Fig. 5a), where a certain HON order is established at finite temperature, there is no finite size dependence for $\mu/\epsilon_H=-0.02$, but the dependence is obvious for $\mu/\epsilon_H=-0.1$.

In order to determine if the van der Waals interaction might affect the values of the critical exponents, we performed some calculations for other values of $\xi$. We did not find any important differences to compare with the $\xi=0$ case. As seen from the values of the critical exponents given in Table I, the transitions to the HON phase at $\xi\neq0$ also belong to the three-state Potts universality class.

\section{Conclusions}

Two peaks in $C_v(T)$ dependences of the BL model at low values of chemical potential might imply that there is some intermediate phase separating the disordered (PM) and HON phases. The intermediate planar phase was found in a similar triangular AFM BC model~\cite{zukovic} and the planar rotator ($p$-state clock)  model~\cite{jose,lapili} at not very large values of $p$. Our analysis of the BL model demonstrates that the high temperature peak of $C_v$ represents the second order phase transition in the three-state Potts universality class and the low-temperature peak is due to Schottky anomaly.

\begin{figure}[] 
\includegraphics[width=0.7\columnwidth]{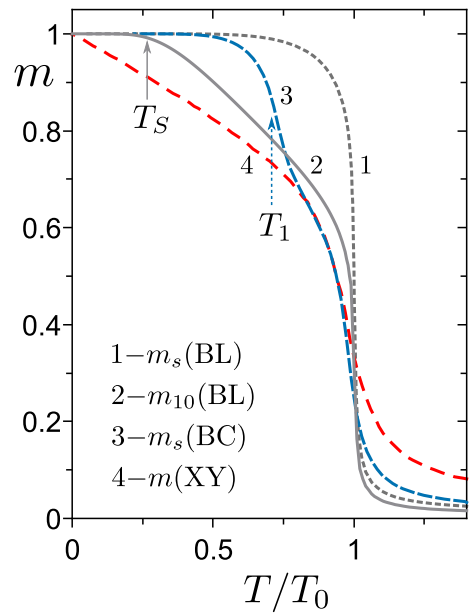} 
\caption{(color online) Temperature dependence of order parameters of the BL model (1 and 2), AFM BC model (3) and $p$-state planar rotator model for $p=128$ (4). The curves 1-3 are obtained for $\mu/\epsilon_H=0.7$ and $L=120$ and the curve 4 - for $L=72$~\cite{lapili}.}  
\label{fig11.eps}
\end{figure} 

In the work of Lapili et al~\cite{lapili} a distinction of phase transitions, characteristic to the Ising-Potts systems, on one hand, and the systems with planar phases, on the other, is given based on the form of the order parameters. Let us analyze the form of temperature dependences of the order parameters in all mentioned models.
In Fig. 11 we present the temperature dependences of the BL model order parameters, $m_s$ and $m_{10}$, as well as those of the triangular AFM BC model (planar phase at intermediate temperatures) and a typical dependence for the planar phase existing all the way down from $T_0=T_2$ (the PM-planar phase transitions point) to $T=0$. The $m_s(T)$ demonstrates a typical Ising-Potts dependence. At the same time, the $m_{10}(T)$ might look similar to a planar phase order parameter. Under scrutiny, the differences are seen. 

For a system, which has the planar phase in between the HON phase at low $T$ and the PM phase at high $T$, the $m_{10}(T)$ should behave in an intricate way similar to that of the AFM BC model (curve 3 in Fig. 11).
Along with the BKT-type transition from the PM to the frustrated phase at $T_0=T_2$ (high $T$), it has to have the frustrated-to-HON transition at $T_1$ (low $T$) and correspondingly demonstrate concaveness of the order parameter at low $T$. We do not find such features in a behavior of the parameter $m_{10}$. At first glance, the $m_{10}(T)$ dependence might look more alike the planar phase order parameter (curve 4) which has the planar phase from $T_2$ down to $T=0$. But differently from the behavior of this order parameter, the $m_{10}(T)$ straightly saturates at the Schottky peak point $T_{S}$.

We did not find the BKT-type phase transition at $T_c$ and further studied the universality class of this transition. In the ground state phase diagram the HON phase is confined ($0<\mu/\epsilon_H<3/2$ at $\xi=0$) between the frustrated phase (similar to that found in the TAFI model) at high particle densities and disordered gas phase at low densities. In contrast to previous MC calculation~\cite{fiore}, which claimed the transition being in the Ising universality class, our calculations demonstrate that the transition at $0.3<\mu/\epsilon_H<1.2$
belongs to the three-state Potts universality class. We determined that the universality of the phase transition changes by approaching both edges of the HON phase. At high densities the critical exponents $\alpha$ and $1/\nu$ systematically decrease from the three-state Potts values (0.33 and 1.2) at $\mu/\epsilon_H=0.5$ towards 0.11 and close to 1, respectively, at $\mu/\epsilon_H=0.05$. Such a behavior implies the approach to crossover from the three-state Potts to the Ising universality class. At another boundary of the HON phase, the critical exponents start to deviate from those of the Potts model at $\mu/\epsilon_H>1.0-1.1$, i. e. closer to the top of the ($T,\mu$) phase diagram in Fig. 3b. At very low particle densities, the transition at $T_c$ is found to be of the first order with the tricritical point being at $\mu/\epsilon_H\approx1.47$. Close to this edge, the behavior is very similar to that found in the triangular AFM BC model.



\begin{thebibliography}{}

\bibitem{bartels}
L. Bartels, Nature Chemistry {\bf 2}, 87 (2010).

\bibitem{barth}
J. V. Barth, Annu. Rev. Phys. Chem. {\bf 58}, 375 (2007). 

\bibitem{dmitriev}
A. Dmitriev, N. Lin, J. Weckesser, J.V. Barth, and K. Kern, J. Phys. Chem. B
{\bf 106}, 6907 (2002).

\bibitem{griessl}
S. Griessl, M. Lackinger, M. Edelwirth, M. Hietschold, and W.M. Heckl, Single
Mol. {\bf 3}, 25 (2002).

\bibitem{li}
Z. Li, B. Han, L.J. Wan, and Th. Wandlowski, Langmuir {\bf 21}, 6915 (2005).

\bibitem{ye}
Y. C. Ye,  W. Sun, Y. F. Wang, X. Shao, X. G. Xu, F. Cheng, J. L. Li, and K. Wu,
J. Phys. Chem. C {\bf 111},  10138 (2007).

\bibitem{kampschulte}
L. Kampschulte, T. L. Werblowsky, R. S. K. Kishore, M. Schmittel, W. M. Heckl,
and M. Lackinger, J. Am. Chem. Soc. {\bf 130}, 8502 (2008).

\bibitem{gutzler}
R. Gutzler, T. Sirtl, J. F. Dienstmaier, K. Mahata, V. M. Heckl, M. Schmittel,
and M. Lackinger, J. Am. Chem. Soc. {\bf 132}, 5084 (2010).

\bibitem{ciesielski}
A. Ciesielski, P. J. Szabelski, W. Rżysko, A. Cadeddu, T. R. Cook, P. J. Stang, and P. Samori,
J. Am. Chem. Soc. {\bf 135}, 6942 (2013).

\bibitem{silly}
F. Silly, A. Q. Shaw, M. R. Castell, G. A. D. Briggs, M. Mura, N. Martsinovich
and L. Kantorovich, J. Chem. Phys. C, 112, 11476, 2008.

\bibitem{mura}
M. Mura, N. Martsinovich and L. Kantorovich, Nanotechnology
19, 465704 (2008).

\bibitem{theobald}
J. A. Theobald, N. S. Oxtoby, M. Phillips, N. R. Champness, P. H. Beton, Nature
{\bf 424}, 1029 (2003).

\bibitem{weber}
U. K. Weber, V. M. Burlakov, L. M. A. Perdigao, R. H. J. Fawcett, P. H. Beton,
N. R. Champness, J. H. Jefferson, G. A. D. Briggs, and D. G. Pettifor, Phys.
Rev. Lett. {\bf 100}, 156101 (2008).

\bibitem{silly1}
F. Silly, U. K. Weber, A. Q. Shaw, V. M. Burlakov, M. R. Castell, G. A. D.
Briggs, and D. G. Pettifor,  Phys. Rev. B, {\bf 77}, 201408 (2008).

\bibitem{perdigao}
L. M. A. Perdig\~ao, G. N. Fontes, B. L. Rogers, N. S. Oxtoby, G. Goretzki, N. R. Champness, and
P. H. Beton, Phys. Rev. {\bf 76}, 245402 (2007).

\bibitem{palma}
C.-A. Palma, J. Bjork, M. Bonini, M. S. Dyer, A. Llanes-Pallas, D. Bonifazi, M. Persson,
and P. Samor\'i, J. Am. Chem. Soc. {\bf 131}, 13062 (2009).

\bibitem{martsi}
N. Martsinovich, A. J. Troisi, Phys. Chem. C {\bf 114}, 4376 (2010).

\bibitem{rochefort}
A. Rochefort and J. D. Wuest, Langmuir, {\bf 25}, 210 (2009).

\bibitem{balbas}
M. Balb\'as Gambra, C. Rohr, K. Gruber, B.A. Hermann, and T. Franosch, Eur. Phys. J. E {\bf 35}, 25 (2012).

\bibitem{misiunas}
T. Misi\={u}nas and E. E. Tornau, J. Phys. Chem. B {\bf 116}, 2472 (2012).

\bibitem{simenas}
M. \v Sim\.enas and E. E. Tornau, J. Chem. Phys. {\bf 139}, 154711 (2013).

\bibitem{simenas1}
M. \v Sim\.enas and E. E. Tornau, J. Chem. Phys. {\bf 141}, 054701 (2014).

\bibitem{wu}
F. Y. Wu, Rev. Mod. Phys. {\bf 54}, 235 (1982).

\bibitem{bc}
M. Blume, Phys. Rev. {\bf 141}, 517 (1966); H. W. Capel, Physica (Utrecht) {\bf 32}, 966 (1966).

\bibitem{bcafm}
A. N. Berker and M. Wortis, Phys. Rev. B 14, 4946 (1976); G. D. Mahan and S. M. Girvin,
Phys. Rev. B {\bf 17}, 4411 (1978).

\bibitem{beg}
M. Blume,  V. J. Emery, and R. B. Griffiths, Phys. Rev. A, {\bf 4}, 1071 (1971);
J. Sivardiere and J. Lajzerowicz, Phys. Rev. A {\bf 11}, 2090, 1975.

\bibitem{wannier}
R. M. F. Houtappel, Physica (Utr.) {\bf 16}, 425 (1950); G. H. Wannier, Phys.
Rev. {\bf 79}, 357 (1950).

\bibitem{baxter}
R. J. Baxter, J. Phys. A {\bf 13}, L61 (1980).


\bibitem{bell}
G. M. Bell and D. A. Lavis, J. Phys. A: Gen. Phys. {\bf 3}, 568 (1970).

\bibitem{barbosa}
M. A. A. Barbosa and V. B. Henriques, Phys. Rev. E {\bf 77} 051204 (2008).

\bibitem{fiore}
C. E. Fiore, M. M. Szortyka, M. C. Barbosa, and V. B. Henriques, J. Chem. Phys. {\bf 131}, 164506 (2009).

\bibitem{MB}
K. A. T. Silverstein, A. D. J. Haymet, and K. A. Dill, J. Am. Chem. Soc. {\bf 120}, 3166 (1998);
T. M. Truskett and K. A. Dill, J. Chem. Phys. {\bf 117}, 5101 (2002); 
J. Phys. Chem. B {\bf 106}, 11829 (2002).

\bibitem{young}
A. P. Young and D. A. Lavis, J. Phys. A: Gen. Phys. {\bf 12}, 229 (1979).

\bibitem{bruscolini}
P. Bruscolini, A. Pelizzola, and L. Casetti, Phys. Rev. Lett. {\bf 88}, 089601 (2002).

\bibitem{patrykiejew}
A. Patrykiejew, O. Pizio, and S. Sokolowski, Phys. Rev. Lett. {\bf 83}, 3442 (1999).

\bibitem{zukovic}
M. \v{Z}ukovi\v{c} and A. Bob\'ak, Phys. Rev. E {\bf 87}, 032121 (2013).

\bibitem{ibenskas}
A. Ibenskas, M. \v Sim\.enas and E. E. Tornau, Phys. Rev. E {\bf 89}, 052144 (2014). 

\bibitem{schick}
M. Schick, J. S. Walker, and M. Wortis, Phys. Rev. B {\bf 16}, 2205 (1977); N.
Berker, S. Ostlund, and F. A. Putnam, Phys. Rev. B {\bf 17}, 3650 (1978);
W. Kinzel, M. Schick Phys. Rev B {\bf 23}, 3435 (1981). 

\bibitem{alexander}
S. Alexander, Phys. Lett. A {\bf 54}, 353 (1975).

\bibitem{blote1}
H. W. J. Bl\"ote and H. J. Hilhorst, J. Phys. A: Math. Gen. {\bf 15}, L631 (1982).

\bibitem{KT}
J.M. Kosterlitz and D.J. Thouless, J. Phys. C: Solid State Phys. {\bf 6}, 1181 (1973).

\bibitem{ono}
I. Ono,  J. Phys. Soc. Jpn. {\bf 53}, 4102 (1984); Prog. Theor. Phys. Supp. {\bf 87}, 102 (1986).

\bibitem{landau}
D. P. Landau, Phys. Rev. B {\bf 27}, 5604 (1983).

\bibitem{blote}
X. Qian and H. W. J. Bl\"ote, Phys. Rev. B {\bf 69}, 036127 (2004);
X. Qian, M. Wegewijs, and H. W. J. Bl\"ote, Phys. Rev. B {\bf 69}, 036127 (2004).

\bibitem{takayama}
H. Takayama, K. Matsumoto, H. Kawahara, and K. Wada, J. Phys. Soc. Jpn. {\bf 52}, 2888 (1983).

\bibitem{miyashita}
S. Miyashita, Proc. Jpn. Acad., Ser. B {\bf 86}, 643 (2010); S. Miyashita, H. Kitatani, and Y. Kanada, J. Phys. Soc. Jpn. {\bf 60}, 1523 (1991).

\bibitem{jose}
J.V. Jos\'e, L.P. Kadanoff, S. Kirkpatrick, and D.R. Nelson, Phys. Rev. B {\bf 16}, 1217 (1977).

\bibitem{tobochnik}
J. Tobochnik, Phys. Rev. B {\bf 26}, 6201 (1982).

\bibitem{noh}
J. D. Noh, H. Rieger, M. Enderle, and K. Knorr, Phys. Rev. E {\bf 66}, 026111 (2002).

\bibitem{cardy}
J. L. Cardy, J. Phys. A: Math. Gen. {\bf 13}, 1507 (1980).

\bibitem{lapili}
C. M. Lapilli, P. Pfeifer, and C. Wexler, Phys. Rev. Lett. {\bf 96}, 140603 (2006).

\bibitem{adler}
J. Adler, A. Brandt, W. Janke, S. J. Shmulyian,  Phys. A {\bf 28}, 5117 (1995).

\bibitem{gorbunov}
V. A. Gorbunov, S. S. Akimenko, A. V. Myshlyavtsev, V. F. Fefelov, M. D. Myshlyavtseva, Adsorption {\bf 19} 571 (2013). 

\bibitem{coddington}
P. D. Coddington and L. Han, Phys. Rev. B {\bf 50}, 3058 (1994).

\bibitem{wang}
Fugao Wang and D. P. Landau, Phys. Rev. Lett. {\bf 86}, 2050 (2001); Phys. Rev E {\bf 64}, 056101 (2001).

\bibitem{reweight}
A. M. Ferrenberg and R. H. Swendsen, Phys. Rev. Lett. {\bf 61}, 2635 (1988);
{\bf 63}, 1658 (1989).

\bibitem{ferrenberg}
A. M. Ferrenberg and D. Landau, Phys. Rev. B {\bf 44}, 5081 (1991).

\bibitem{challa} 
M. S. S. Challa, D. P. Landau, and K. Binder, Phys. Rev. B {\bf 34}, 1841
(1986).

\bibitem{tome}
T. Tome and A. Petri, J. Phys. A: Math. Gen. {\bf 35}, 5379 (2002).

\end{thebibliography}
\end{document}